\documentclass[amsmat,amssymb,amsfonts,aps,prb,twocolumn,showpacs]{revtex4}

\usepackage{amsmath}
\usepackage{graphicx}
\usepackage{hyperref}
\usepackage{blkarray}

\usepackage{xcolor}

\usepackage{soul}

\begin{document}

\title{Interplay between interlayer exchange  and stacking in CrI$_3$ bilayers}

\author{D. Soriano$^{\rm{1,2}}$ ,
        C. Cardoso$^{\rm{1,3}}$ ,
        J. Fern\'andez-Rossier$^{\rm{1,4}}$
}

\affiliation{$^{\rm{1}}$QuantaLab, International Iberian Nanotechnology Laboratory (INL), Av. Mestre Jos\'e Veiga, 4715-330 Braga, Portugal} 
\affiliation{$^{\rm{2}}$Radboud University, Institute for Molecules and Materials, NL-6525 AJ Nijmegen, the Netherlands}
\affiliation{$^{\rm{3}}$CNR-Nanoscience Institute, S3 Center, 41125 Modena, Italy}
\affiliation{$^{\rm{4}}$Departamento de F{\'i}sica Aplicada, Universidad de Alicante, 03690, Spain}

\begin{abstract}

We address the interplay between stacking and interlayer exchange  for  ferromagnetically ordered CrI$_3$, both for bilayers and bulk.
Whereas bulk CrI$_3$ is ferromagnetic, both  magneto-optical and transport experiments show that interlayer exchange for CrI$_3$  bilayers is antiferromagnetic.  
Bulk CrI$_3$ is known to assume two crystal structures,  rhombohedral and  monoclinic, that differ mostly in the stacking between  monolayers.  Below 210-220 Kelvin, bulk CrI$_3$  orders in a rhombohedral phase.  Our density functional theory calculations show a very strong dependence of interlayer exchange and stacking.  Specifically,   the ground states of both bulk and free-standing CrI$_3$ bilayers are ferromagnetic for the rhombohedral phase. In contrast, the energy difference between both configurations is more than one order of magnitude smaller for the monoclinic phase, and eventually becomes antiferromagnetic when either  positive strain or on-site Hubbard interactions ($U \geq 3$) are considered. We also explore the interplay between interlayer hybrydization and stacking,  using a Wannier basis, and between interlayer hybrydization and relative magnetic alignment for CrI$_3$ bilayers, that helps to account for the very large tunnel magnetoresistance obvserved in recent experiments.             
\end{abstract}

\maketitle


\section{Introduction} 
The recent discovery  of several  2D ferromagnetic materials \cite{GongZhang2017,HuangXu2017,McGuire2017,FeiXu2018,OHaraKawakami2018} is expanding the research horizons in 2D Materials. These new materials, and more specifically CrI$_3$, open new venues for the fabrication of low dimensional spintronic\cite{SongXu2018,KleinJarillo2018,WangMorpurgo2018} and optoelectronic\cite{SeylerXu2018,ZhongXu2017,HuangXu2018,JiangMak2018,JiangShan2018} devices based on multi-layer structures, and are being the object of strong interest \cite{LadoRossier2017,LiuAnantram2018,LiuPetrovic2018,RichterKlaui2018,ZhangYang2018,ZollnerFabian2017, jiang2018spin,liu2018electrical,CardosoRossier2018,TongYao2018}. Importantly, some of these applications rely on the antiparallel interlayer alignment at zero field, which can be reverted by the application of a magnetic field  \cite{SongXu2018,KleinJarillo2018,WangMorpurgo2018} or, intriguingly, electric\cite{HuangXu2018,JiangMak2018,JiangShan2018} fields.

The family of CrX$_3$ (X = Cl, Br, I) magnetic insulators is representative of this type of 2D ferromagnetic materials. In the single layer limit, magnetic order is very sensitive to magnetic anisotropy, that is governed by anisotropic superexchange in the case of CrI$_3$.\cite{LadoRossier2017,LeeHammel2019,BelbesSolovyev2019,XuBellaiche2018} Experiments in bulk show that CrCl$_3$ is the only one showing an anti-ferromagnetic (AF) order\cite{MorosinNarath1964}, while CrBr$_3$ and CrI$_3$ are bulk ferromagnets\cite{HandyGregory1952,McGuireSales2015} with Curie temperatures $T_c = 37$ and $61$ K, respectively. 
In contrast with bulk,  interlayer coupling for few layer  CrI$_3$ is found to be antiferromagnetic,  based on optical\cite{HuangXu2017,HuangXu2018,JiangMak2018,JiangShan2018}, transport \cite{KleinJarillo2018,WangMorpurgo2018,SongXu2018,KleinJarillo2019} and, more recently, microscopic probes.\cite{ThielMaletinsky2019} This provides a first motivation for  this work.

The second motivation arises from the following observation. Bulk CrI$_3$ undergoes a structural transition\cite{McGuireSales2015} at 210-220 Kelvin, between a  rhombohedral phase at low temperature  and monoclinic  phase at higher temperature. The layer stacking in these structures is shown in Figure \ref{fig1}(a,b). Interestingly,  the differential magnetic susceptibility, $\frac{d\chi}{dT}$, presents a kink  at the structural transition\cite{McGuireSales2015}, which  is consistent with a variation of the interlayer exchange interaction.

In monolayer CrI$_3$, the intralayer FM coupling, that ultimately drives the long-range ordering between Cr atoms, can be anticipated by the Goodenough-Kanamori rules\cite{Goodenough1955,Goodenough1958,Kanamori1959} of single-ligand superexchange (M-L-M), on account of the almost perpendicular alignment between the Cr-I-Cr bonds. The extension of these rules for more than one ligand, in order to predict the interlayer exchange coupling in van der Waals structures (M-L---L-M), is not straightforward.\cite{FeldkemperWeber1998} In the following, we assume a different approach aiming to address the interplay between stacking and interlayer exchange coupling for CrI$_3$ bilayer, combining density functional theory and an effective interlayer coupling model.

\section{Methodology}
\begin{figure}[t]
    \centering
    \includegraphics[clip=true, width=0.45\textwidth] {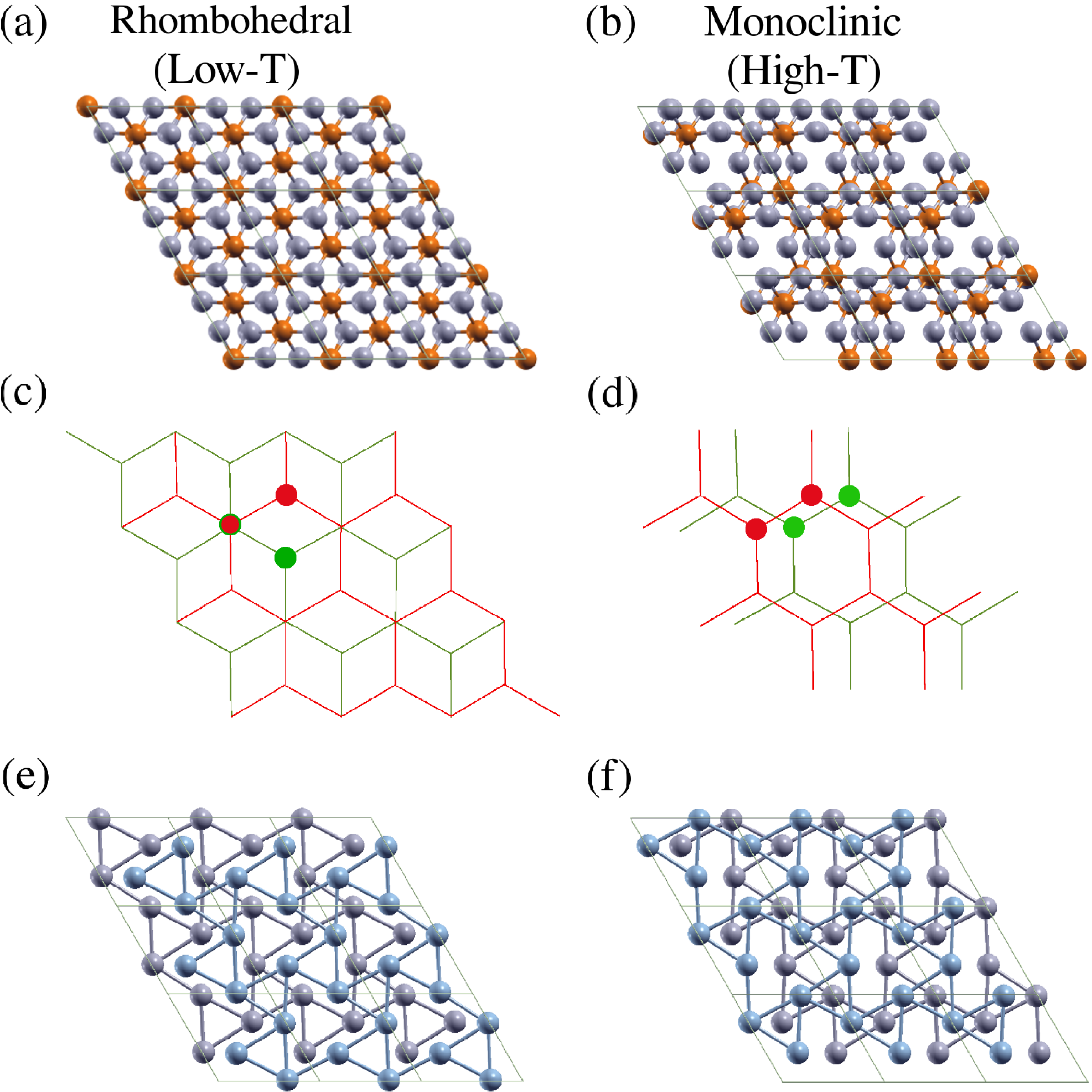}
    \caption{\label{fig1}(Color online) Atomic structure of bilayer CrI$_3$ with (a) rhombohedral and (b) monoclinic crystal structure. These structures are found in bulk CrI$_3$ at low and high temperature. Orange and gray atoms correspond to Cr and I, respectively. (c) and (d) show the AB and AA$_{1/3}$ stacking of the hexagonal Cr lattices between the layers. The green and red lines correspond to the top and bottom hexagonal Cr lattices respectively.  The green and red dots denote the Cr atoms in the top and bottom layers in  the elementary unit cell. (e) Detail of the atomic structure of the I atoms at the bilayer interface for the rhombohedral stacking. (f) Idem for the monoclinic one. }
\end{figure}

 Our calculations are based on density functional theory. For each CrI$_3$ stacking shown in Fig.\ref{fig1}(a), we first perform a geometry relaxation starting from previously reported experimental crystal structure ($a = b = 6.867$ \AA~  for the rhombohedral structure and $6.866$ \AA~  for the monoclinic one).\cite{McGuireSales2015} The relaxation is carried out using the plane-wave based code PWscf as implemented in the Quantum-Espresso \textit{ab-initio} package\cite{QE}. For the self consistent calculations, we use a $8 \times 8 \times 1$ $k$-point grid for the bilayer calculations and a $12 \times 12 \times 6$ k-mesh for the bulk.  Projector augmented wave (PAW) pseudopotentials and the Perdew-Burke-Ernzerhoff (PBE) approximation\cite{perdew96} for the exchange-correlation functional are used for Cr and I atoms. Van der Waals interactions are included through the Grimme-D2 model.\cite{grimme06} Spin-orbit coupling is not considered in these calculations.

\section{Interlayer exchange and stacking} 
 Figures \ref{fig1}(c) and \ref{fig1}(d) show the two Cr hexagonal lattices (in red and green) and the stacking details of the bilayer. In the rhombohedral case, the two Cr lattices follows an AB or Bernal stacking, similar to bilayer graphene. The monoclinic case can be obtained by starting with an AA stacking and displacing the top layer $a$/3 along one of the in-plane lattice vectors $\vec{a}$ or $\vec{b}$ (we have labeled this stacking as AA$_{1/3}$). The two atomic structures induce a different arrangement of the I atoms at the interface which is shown in Figure \ref{fig1}(e,f). The position of the I atoms has an important effect on the interlayer coupling since it affects the Cr-Cr interlayer distance through steric effects (see Table \ref{tab1}).

We have performed first-principles calculations of bilayer (bulk) CrI$_3$ using both structures, namely AB (rhombohedral) and AA$_{1/3}$ (monoclinic), with different interlayer magnetic order (FM \textit{vs} AF). Our results are summarized in table (\ref{tab1}). We find that, for both bilayer and bulk CrI$_3$ the energy difference between the FM and AF configurations is dramatically reduced in the case of monoclinic stacking. As we discuss in section \ref{pert},  two different perturbations lead to an antiferromagnetic interlayer interaction  for the $AA_{1/3}$ stacking of the bilayer: addition of a Hubbard $U$ correction, keeping the same geometry obtained without U, and modification of the interlayer distance.

\begin{table*}[t]
\centering
\caption{Summary of the calculations for bilayer (AA$_{1/3}$ \emph{vs} AB) and bulk (Monoclinic \emph{vs} Rhombohedral) CrI$_3$. The equilibrium interlayer distance ($d_{in}^{\rm PW}$) is obtained by relaxing the geometry starting from the experimental crystal structure\cite{McGuireSales2015}. We keep the same geometry for the two functionals, namely PBE-D2 and PBE+U-D2. D2 stands for the Grimme-D2 van der Waals correction\cite{grimme06}. The energy difference is defined as $\Delta E = E_{AF} - E_{FM}$. The values of the exchange interlayer coupling are obtained from Equations \ref{bilayer} and \ref{bulk} for bilayer and bulk respectively.  The last column shows the values of $\overline{j}_{\rm 12}/N_{\rm at}$, where $N_{\rm at}$ is the number of atoms in the unit cell.}  
\label{tab1}
\begin{tabular}{|l|c|c|c|c|}
\hline
& $d_{in}^{\rm PW}$ (\AA) & $\Delta E^{\rm PW}$ (meV) & $\overline{j}_{\rm in}$ ($\mu$eV) & $\overline{j}_{\rm in}$ ($\mu$eV/u.c.)   \\
\hline \hline
Bilayer AA$_{1/3}$ (PBE-D2)   & 6.621 & 0.21  & 46.7   & 11.7     \\
Bilayer AB (PBE-D2)           & 6.602 & 9.43  & 2095.6 & 523.9    \\
Bilayer AA$_{1/3}$ (PBE+U-D2) & 6.621 & -0.36  &  -80.7  &  -20.2    \\
Bilayer AB (PBE+U-D2)         & 6.602 & 17.82  & 3959.7  & 989.9  \\
\hline
Bulk Monoclinic (PBE-D2)      & 6.621 & 0.44  & 16.3   & 1.4      \\ 
Bulk Rhombohedral (PBE-D2)    & 6.602 & 4.50  & 166.7  & 13.9     \\
\hline
\end{tabular} 
\end{table*}

We now discuss how to relate the DFT results to the average interlayer exchange coupling ($\overline{J}_{12}$). For that matter, we assume that the interlayer exchange can be described by a classical Heisenberg model:
\begin{equation}
{\cal H}_{\rm inter}= -\sum_{i\in 1, j\in 2} J^{\rm inter}_{ij} \vec{S}_i\cdot\vec{S}_j
\end{equation}
where $1$ and $2$ label the two CrI$_3$ layers and $J_{ij}$ is are the interlayer exchange interactions.  The sign convention we take is such that $J_{ij}>0$  ($J_{ij}<0$) stands for ferromagnetic (antiferromagnetic) interaction.
Assuming that all spins are parallel, and have a  length $S$,  the energy for configurations where all spins in a given layer are parallel, and collinear with those of the other layer,  is   $U=\mp S^2 \overline{J}_{12}$ where
\begin{equation}
\overline{J}_{12} =\sum_{i\in {1}, j\in {2}} J^{\rm inter}_{ij}
\end{equation}
is the {\em average interlayer coupling} 
 and the sign  $-$ ($+$)   corresponds to the ferromagnetic (antiferromagnetic)  interlayer alignment.  It is self-evident that $\overline{J}_{12} $ is an increasing  function of the number of atoms in the unit cell.  For CrI$_3$, there are two atoms per unit cell and plane. 

We now break down this average exchange, and the corresponding total interlayer exchange, as a sum over the contribution coming from each unit cell ,  $\overline{J}_{12} =N \overline{j}_{12} $, where $N$ is the number of unit cells and $\overline{j}_{12} $ is split as the sum of intracell and intercell contributions.
\begin{equation}
\overline{j}_{12} =\sum_{i\in (I,1), j\in (I,2)} J^{\rm intra}_{ij} + \sum_{\in(I,1) ,j  \in (I',2)} J^{\rm inter}_{ij}
\end{equation}
Since $J^{\rm inter}_{ij}$ decays very rapidly with distance,  $\overline{j}_{12}$ converges.  Therefore, in the case of the bilayer,  the total energy per unit cell, that can be compared with DFT calculations, reads as an Ising model for a dimer:

\begin{equation}
{\cal U}_{\rm in}(s_1,s_2)= -\overline{j}_{12}  S^2  s_{1}s_{2}
\label{eqn1}
\end{equation}
where $s_1,s_2=\pm 1$ describe the orientation of the layer magnetization.  As a result, we can relate the energy difference between the parallel and antiparallel configurations in the DFT calculations with the average interlayer exchange, through 
\begin{equation}
\Delta E\equiv {\cal U}_{\rm in}(+,-)-{\cal U}_{\rm in} (++)= 2 \overline{j}_{12}S^2
\label{bilayer}
\end{equation}

We now carry out the same analysis for   the case of bulk, the unit cell has 3 planes.  Therefore,  the effective model has to keep track of the magnetization of 6 layers:
\begin{equation}
{\cal U}_{\rm in}(s_1,s_2,s_3,s_4,s_5,s_6)= -\overline{j}_{12}  S^2  \sum_{i=1,6} s_{i}s_{i+1}
\label{eqn1bulk}
\end{equation}
where we assume $s_1=s_7$, to account for the periodic boundary conditions in the off-plane direction. 
Thus, for the bulk calculations we have:
\begin{equation}
\Delta E\equiv  {\cal U}_{\rm in}(+-+-+-)-{\cal U}_{\rm in} (++++++)= 12 \overline{j}_{12}S^2
\label{bulk}
\end{equation}

Equations  (\ref{bilayer}) and  (\ref{bulk}) permit  to relate our DFT calculations with the average interlayer exchange.   By so doing, we find that the interlayer exchange shows always a stronger ferromagnetic character than the AB (rhombohedral) phase (see Table \ref{tab1}) than the AA$_{1/3}$ (monoclinic). This clearly indicates that there is a correlation between stacking geometry and the interlayer exchange. This is expected,  since different stacking  imply both different interlayer  Cr-Cr distances and  Cr-I bond angles (see figure \ref{fig1}), which are the structural variables that control exchange.

  \section{Relation with experiments}
  
Our results for bulk are consistent with the ferromagnetic interlayer interaction observed experimentally. In addition,  our results  might help to understand  the kink in the magnetic susceptibility observed by McGuire \textit{et al.} \cite{McGuireSales2015} at the structural  phase transition observed in bulk CrI$_3$ at 220 Kelvin, between a low temperature rhombohedral and a high temperature monoclinic structures. The connection can be established as follows. At high temperature,  CrI$_3$ is paramagnetic. The  susceptibility of a ferromagnet in the paramagnetic regime is described by the Curie law,
\begin{equation}
\chi_{CW}
=
\frac{S(S+1) (g\mu_B)^2 }{3k_B }\frac{1}{T-T_{CW}}
\label{eqn4}
\end{equation}
where $S$ is the total spin of the Cr atoms, $g$ is the g-factor, $\mu_B$ is the Bohr magneton, $k_B$ is the Boltzmann constant and $T_{CW}$ is the Curie temperature, that logically depends on both the interlayer and intralayer couplings through the relation 
\begin{equation}
3k_B T_{CW}= S(S+1)\overline{J}
\label{eqn5}
\end{equation}
where $\overline{J} = \sum_{j} (J^{\rm inter}_{ij}+J^{\rm intra}_{ij})$ is the sum of all the exchange interactions for a given spin $i$.  We are assuming here that the average magnetization of all spins is the same.  The calculated variation  of the interlayer coupling for the two different stackings, shown in the table,  will lead to a modification of $T_{CW}$ ,  $\Delta T_{CW}$ at the temperature of the structural transition,  that leads to an additional contribution to $\frac{d\chi}{dT}\propto \frac{\partial \chi}{\partial T_{CW}} \Delta T_{CW}$.     

We now discuss the relation of our results with experimental results for CrI$_3$ bilayers and thin films\cite{HuangXu2017,HuangXu2018,JiangMak2018,JiangShan2018,KleinJarillo2018,WangMorpurgo2018,SongXu2018,ThielMaletinsky2019}, showing antiferromagnetic interlayer interaction. The application of off-plane magnetic fields of $B_c\simeq 0.35 T$ and $0.6 T$ revert the interlayer magnetization in CrI$_3$ bilayers as recently proven by transport\cite{KleinJarillo2018} and optical\cite{HuangXu2017} measurements respectively.  We can estimate the interlayer exchange by equating the Zeeman energy per unit cell, $E_Z= 4\times g\mu_B  S B_c$ to $\Delta E$ in eq. (\ref{eqn1}).  We thus obtain  $\overline{j}_{12}= \frac{-2g}{S}\mu_B B_c\simeq -60\mu eV $ and $-103\mu eV$.

Our DFT results show that interlayer exchange is much smaller  for the AA$_{1/3}$ stacking, although still weakly ferromagnetic.   Other density functional calculations, appeared after a first version of our work was posted in the arXiv,    show that interlayer exchange can indeed become antiferromagnetic for the  AA$_{1/3}$stacking,  using functionals different from GGA\cite{WangMorpurgo2018,JiangJi2018,SivadasXiao2018,JangHan2018,LeiMacDonalds2019}. The common point in all DFT calculations is that interlayer exchange has a {\em weaker ferromagnetic character} for the AA$_{1/3}$ stacking than for the AB.

Our DFT calculations still predict that  the AB stacking is the  ground state structure for the freestanding bilayer.  However,  the energy difference between these two stacking configurations is $\Delta E_{\rm struct} \approx 3.25$ meV/Cr atom, much smaller than its bulk counterpart,  $\Delta E_{\rm struct} \approx 8$ meV/Cr atom favouring the rombohedral crystal structure. Given that experiments are always carried out with the CrI$_3$ bilayers deposited on top of substrates, such as graphene and silicon oxide,  it can be that these favour the AA$_{1/3}$ stacking, leading to an antiferromagnetic interaction. It is also possible that stacking energetics is different in bulk and in very thin films, on account of the different contributions coming from long-range dispersive forces in both cases. Recent  experimental work \cite{ThielMaletinsky2019} provides evidence that this might be indeed the case. 

\section{Effect of interlayer distance and on-site Hubbard interaction ($U$) on Interlayer exchange
\label{pert}}We now consider two types of  perturbations that could further reduce  the ferromagnetic interlayer exchange  and eventually yield an  antiferromagnetic coupling, namely  a modification of the interlayer distance and the inclusion of an on-site Hubbard interaction ($U$) using the so called PBE+U functional, in the spirit of the LDA+U approximation\cite{anisimov91}. The first one could be driven by the coupling to the substrate, whose effect is missing in our DFT calculations.  For instance, charge transfer is predicted to occur in the graphene/CrI$_3$ interface\cite{CardosoRossier2018}, that could modify interlayer separation. 

\begin{figure}[t]
    \centering
   \includegraphics[clip=true, width=0.4\textwidth] {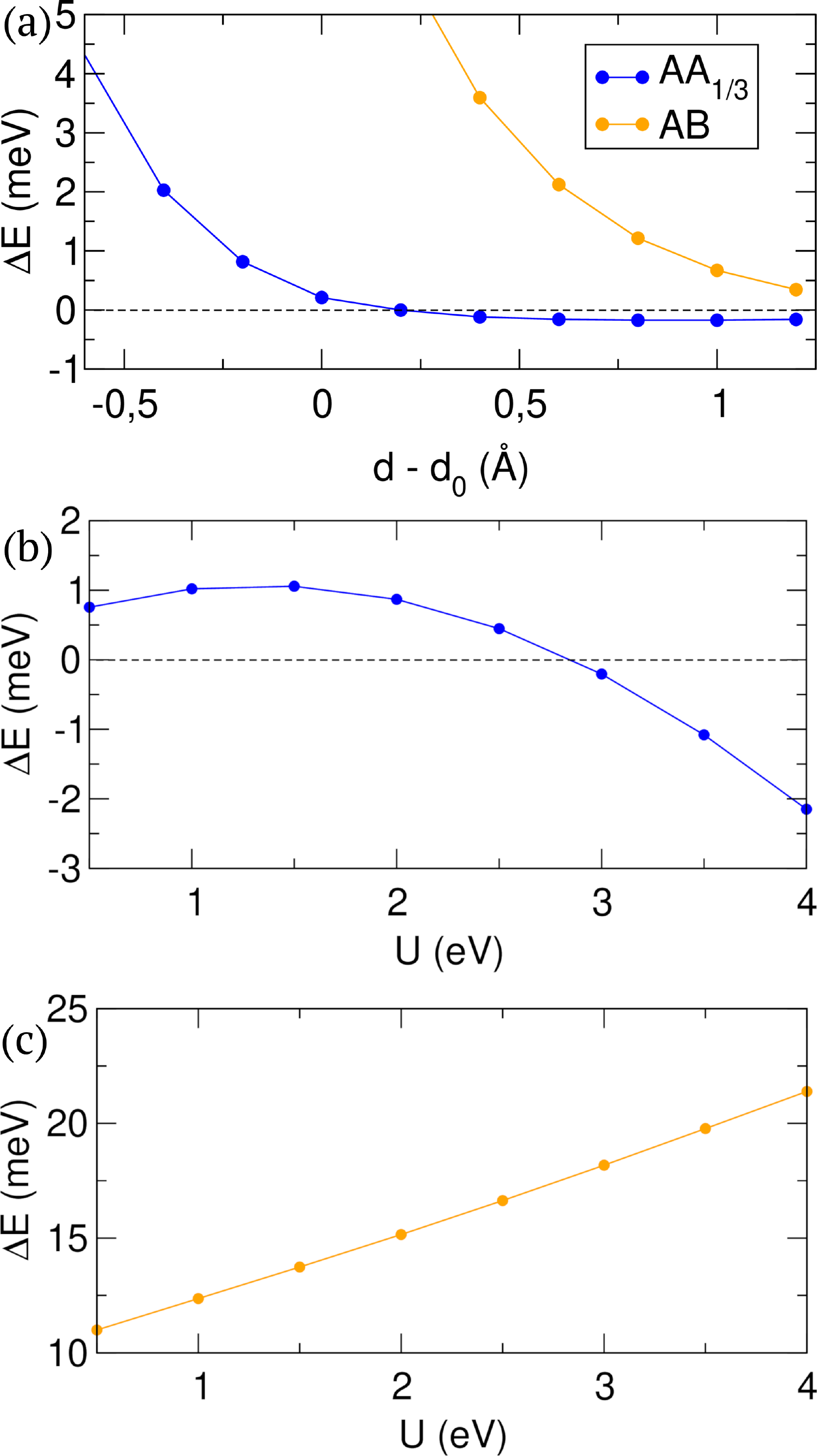}
    \caption{\label{fig2}(Color online) (a) Energy difference $\Delta E = E_{AF} - E_{FM}$ of bilayer CrI$_3$ for AB (orange) and AA$_{1/3}$ (blue) stacking. $\Delta E < 0$ stands for AF phase transition (horizontal dashed line). For $d - d_0 > 0.2$ \AA, the AA$_{1/3}$ bilayer becomes AF. (b) Scaling of $\Delta E$  for different values of the Coulomb repulsion ($U$) in AA$_{1/3}$ stacking. For $U \geq 3$ the AA$_{1/3}$ (monoclinic) stacking becomes AF. (c) Idem for the AB (rhombohedral) stacking. In contrast to the AA$_{1/3}$ case, the AB case never becomes AF when increasin the Coulomb interaction.    }
\end{figure}

Figure \ref{fig2}(a) shows the energy difference $\Delta E = E_{AF} - E_{FM}$ for different interlayer distances and for both structures. For $\Delta E < 0$ the interlayer exchange coupling becomes AF (horizontal dashed line). This occurs in the AA$_{1/3}$ stacking for $d - d_0 > 0.2$~ \AA, while the AB CrI$_3$ remains FM.
The maximal value for the AF exchange, $j_{AF}= -38 \mu eV$ is obtained for $d-d_0 = 1$~ \AA.

We now discuss the scaling of $\Delta E$ as we change the on-site Hubbard inderaction $U$, keeping the same geometry obtained for $U=0$. Our results are shown in  
figure \ref{fig2}(c,d),  for the AA$_{1/3}$ and AB stacking respectively. We observe that, in contrast to the AB stacking, the interlayer exchange in the AA$_{1/3}$ case scales non-monotonically with $U$ and, for $U \geq 3$ it becomes AF (dashed line indicates the transition from positive to negative exchange coupling).

 Interestingly, we find that the response of the system to  both the structural modification and the addition of a Hubbard $U$ correction, follows the same pattern. First,  none of these perturbations drive the system to the AF interlayer interaction in the case of the AB stacking. Second, both perturbations drive the interlayer interaction AF in the $AA_{1/3}$ stacking. Third, the dependence of $\Delta E$ on both interlayer distance and $U$ is non-monotonic only in the case of the $AA_{1/3}$ stacking.


Given that all known mechanisms for exchange lead to monotonic dependence with distance, the non-monotonic behaviour of $\Delta E$, for the AA$_{1/3}$ stacking, that includes a change of sign,   clearly shows that interlayer exchange interaction is the result of at least two  contributions with opposite signs:
\begin{equation}
J_{\rm in}= J^{\rm FM}_{\rm in} - J^{\rm AF}_{\rm in}.
\label{eqn6}
\end{equation}
The first contribution, which favours a ferromagnetic coupling,  arises both  from interlayer superexchange pathways and direct exchange.  The second contribution favours antiferromagnetic exchange.

\section{Interplay between interlayer hybridization and stacking}
We now explore if interlayer antiferromagnetic exchange could be accounted for by  the theory of kinetic exchange of Anderson\cite{Anderson1963} (see also Hay \textit{et al.}\cite{HayHoffman1975}), for electrons occupying otherwise degenerate orbitals that become weakly hybridized by an interlayer hopping $\gamma$. The interlayer hopping leads to the formation of  bonding-antibonding states that delocalize the states among the two layers.  In the limit of   on-site Hubbard repulsion $U$ much larger than $\gamma$,  the low energy states of this  Hubbard dimer are described by a spin Heisenberg model with  antiferromagnetic exchange $J^{\rm AF}_{\rm in}=\frac{4\gamma^2}{\tilde{U}}$. Here, we use a Hubbard $\tilde{U}$ to differentiate it from the Hubbard $U$ used in LDA+U calculations. The former it is always present even for $U=0$ calculations, and stands for the energy that should be paid to doubly occupy an atomic orbital. The second one is the extra energy that should be paid when the orbitals are strongly localized in order to account for correlation effects.

\begin{figure}[t]
    \centering
   \includegraphics[clip=true, width=0.5\textwidth] {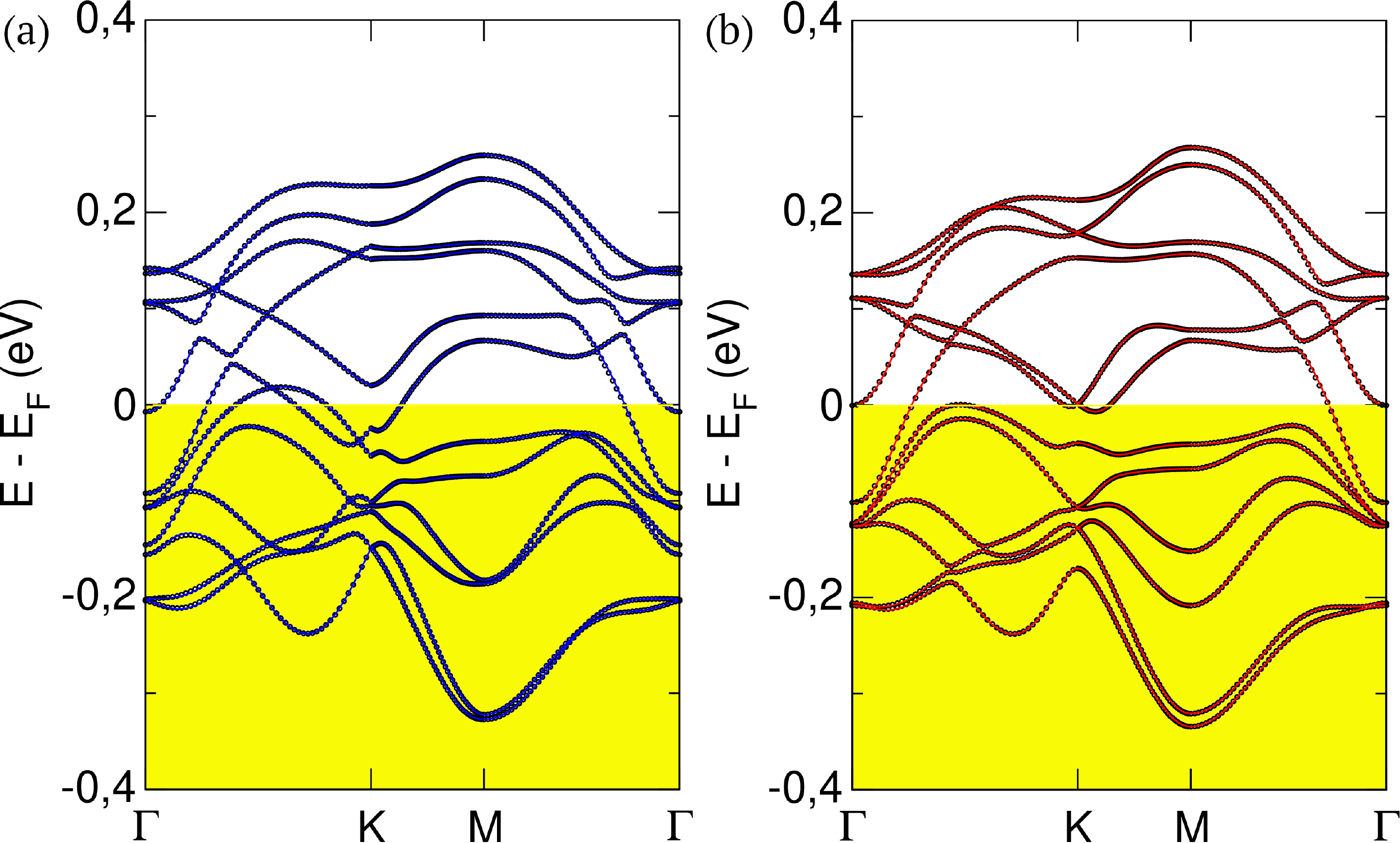}
    \caption{\label{fig3}(Color online) Spin unpolarized band structures of bilayer CrI$_3$ for (a) AA$_{1/3}$ and (b) AB stacking configurations. The half-filled bands correspond to the $t_{2g}$ bands of Cr. The red and blue lines on top of the $t_{2g}$ bands corresponds to the Wannier bands. 
    }
\end{figure}

In order to   explore whether the interlayer hybdridization $\gamma$ is significantly different for the two stacking geometries for CrI$_3$ bilayer,  we calculate the hybridization between crystal-field split $t_{2g}$.   For that matter we obtain a tight-binding model from our DFT calculations,  using a representation of the DFT hamiltonian in a basis of maximally localized Wannier orbitals.    To do so,  we use DFT as implemented in the plane-wave based \emph{PWscf} code (see \emph{Methodology} section),  with  {\em spin-unpolarized solutions},   to ensure that the band splitting comes only from the interlayer hybridization. In the non-magnetic  solutions, the  $t_{2g}$ bands are  half filled, in contrast to the spin-polarized case, where the 3 electrons with same spin fill the 3 $t_{2g}$ bands in one spin channel.  The spin-unpolarized $t_{2g}$ bands of bilayer CrI$_3$ for AA$_{1/3}$ and AB cases are shown in Figure \ref{fig3}(a,b).

In order to obtain a representation of the Hamiltonian in a basis of atomic-like orbitals,  we transform our plane-wave basis into a localized Wannier one using \emph{wannier90} code\cite{Wannier90}. The representation of the Hamiltonian in that basis  allows us to extract the interlayer hopping amplitudes $\Gamma_{ij}$ from the Wannier Hamiltonian. We choose a projection over the subspace spanned by the $t_{2g}$ manifold, namely $\{d_{xy},d_{xz},d_{yz}\}$ centered in the Cr atoms. Red and blue lines on top of the $t_{2g}$ bands in Figure \ref{fig3}(a,b) correspond to the Wannier bands obtained for each stacking configuration. The Wannier Hamiltonian for intracell atoms takes the form 
\begin{equation}
H_W = 
\begin{pmatrix} 
E^1_{t_{2g}} & \Gamma^{12} \\
\Gamma^{21} & E^2_{t_{2g}}
\end{pmatrix}
\label{eqn7}
\end{equation}
where $E^1$ and $E^2$ are 6$\times$6 matrices containing the on-site energies of the $t_{2g}$ orbitals in layer $1$ and layer $2$. The $\Gamma$ matrices contain the hopping terms connecting both layers. Equations \ref{eqn8} and \ref{eqn9} show the detailed structure of the coupling matrices. The ball and stick models close to the matrices indicate the unit cell atoms for both stacking configurations. Atoms with different colors belong to different layers.  
\begin{equation}
\Gamma_{AB}  =   
\begin{blockarray}{ccccccc}
        & d^3_{xy} & d^3_{xz} & d^3_{yz} & d^4_{xy} & d^4_{xz} & d^4_{yz}\\
      \begin{block}{c(cccccc)}
         d^1_{xy}~ & 0 & 0 & 0 & 0 & 0 & 0 \\
         d^1_{xz}~ & 0 & -22 & 0 & 0 & 0 & 0 \\
         d^1_{yz}~ & 0 & 0 & 0 & 0 & 0 & -29 \\
         d^2_{xy}~ & 0 & 22 & -21 & 0 & 0 & 0 \\
         d^2_{xz}~ & 22 & 0 & 22 & 0 & 0 & 0 \\
         d^2_{yz}~ & -21 & 22 & 0 & 0 & 0 & -22 \\
      \end{block}
    \end{blockarray}
    \qquad
\raisebox{0mm}{\includegraphics[keepaspectratio = true, scale = 0.25] {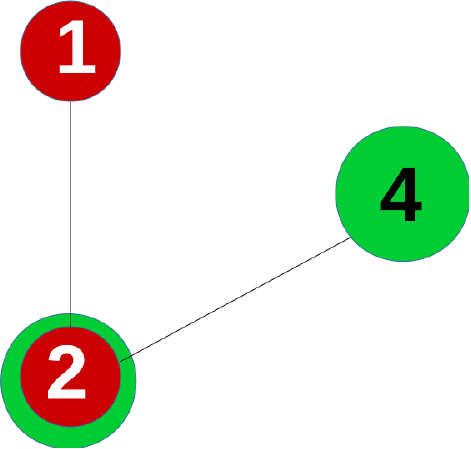}}
\label{eqn8}      
\end{equation}
\begin{equation}
\Gamma_{AA}  =  
\begin{blockarray}{ccccccc}
        & d^3_{xy} & d^3_{xz} & d^3_{yz} & d^4_{xy} & d^4_{xz} & d^4_{yz}\\
      \begin{block}{c(cccccc)}
         d^1_{xy}~ & 0 & 29 & 0 & 0 & 0 & -25 \\
         d^1_{xz}~ & 23 & 0 & 0 & 0 & 0 & 0 \\
         d^1_{yz}~ & 0 & 0 & 0 & -25 & 0 & 0 \\
         d^2_{xy}~ & 0 & 0 & 0 & 0 & 23 & 0 \\
         d^2_{xz}~ & 0 & 0 & 0 & 29 & 0 & 0 \\
         d^2_{yz}~ & 0 & 0 & 0 & 0 & 0 & 0 \\
      \end{block}
    \end{blockarray}    
\qquad
\raisebox{0mm}{\includegraphics[keepaspectratio = true, scale = 0.25] {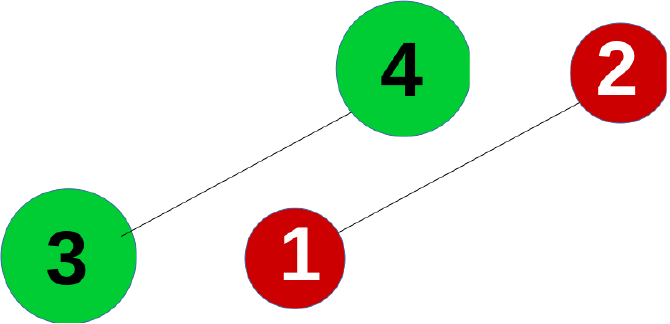}}    
\label{eqn9}
\end{equation}

Inspection of  the elements of the $\Gamma$ matrices for each stacking configuration,  show that each Cr atom is connected at least with two Cr atoms.  Thus, interlayer exchange couples a given Cr atom in a layer with several Cr atoms in the other layer.   Also, the directionality of $d$-type orbitals together with the contribution of the iodine atoms at the interface makes difficult to compare this system with a typical single-orbital based model of bilayer honeycomb lattice. Comparing both hopping matrices, we observe that the higher contribution to the antiferromagnetic kinetic exchange in the AB case comes only from the interaction between $d_{yz}$ orbitals in atoms 1 and 4 ($\Gamma^{14}_{AB} = -29$ meV). In contrast, the AA$_{1/3}$ interlayer hopping matrix shows two important contributions ($\Gamma^{13}_{AA} = \Gamma^{24}_{AA} = 29$ meV) between orbitals $d_{xy}$ and $d_{xz}$. However, from this analysis, we can not conclude that the average  interlayer hybridization is very different for the two stackings.  Therefore,  the mechanism that accounts for the different interlayer exchange  interaction must arise from other exchange mechanism. 

\section{ Spin polarized energy bands and implications for vertical transport}

\begin{figure}[t]
    \centering
    \includegraphics[clip=true, width=0.45\textwidth] {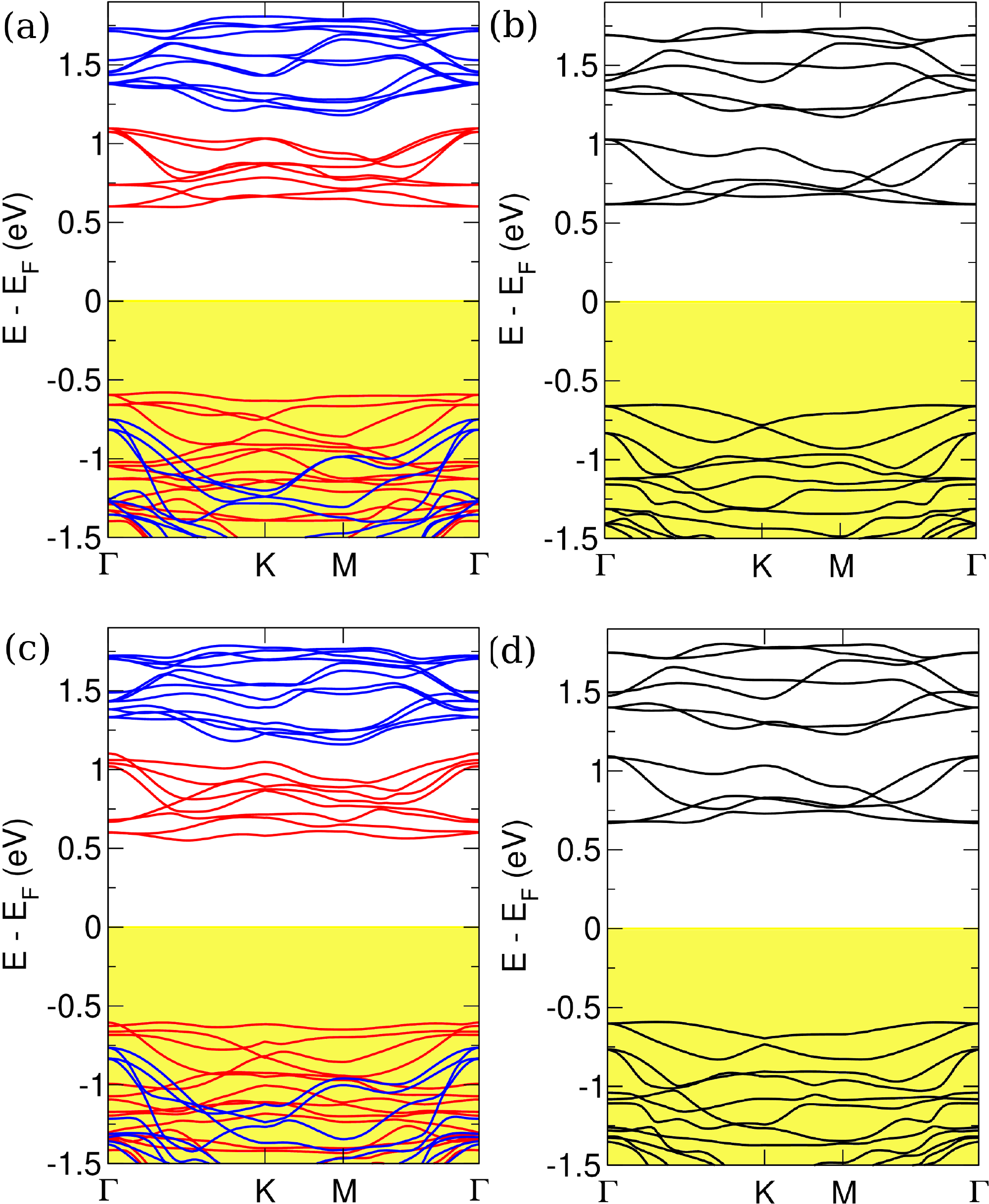}
    \caption{\label{fig4}(Color online) Spin-polarized band structures of bilayer CrI$_3$ for (a) AB(FM), (b) AB(AF), (c) AA$_{1/3}$(FM) and (d) AA$_{1/3}$(AF).  
    }
\end{figure}

We now discuss the spin-polarized band structures obtained from first-principles calculations. In Figure \ref{fig4}, we show the spin polarized bands of the AB (top panels) and AA$_{1/3}$ (bottom panels) bilayer CrI$_3$. For the spin polarized calculations, the spin majority $t_{2g}$ bands are fully occupied and the first set of empty bands is made of spin majority $e_g$ states. The FM cases (a,c) show a clear band splitting of the majority $e_g$ (red bands above the Fermi energy) and minority $t_{2g}$ bands (blue bands above the Fermi energy) coming from the interlayer coupling. This is similar to the unpolarized calculations in Figure \ref{fig3}. 

In contrast, in the antiparallel interlayer (AF)  cases (b,d), the interlayer splitting is absent. The reason is that $e_g$ and $t_{2g}$ for a given spin channel in one layer are degenerate with the same bands with {\em opposite spin} in the other layers.  Since interlayer coupling is spin conserving,  the resulting hybridization is dramatically reduced.  This difference of  interlayer coupling  in the FM and AF configurations  definitely contributes to explain the very large magnetoresistance observed in vertical transport with CrI$_3$ bilayers in the barriers.\cite{KleinJarillo2018,WangMorpurgo2018,SongXu2018}  
Given that the $e_g$ states are the lowest energy channels available for tunneling electrons in the barrier,  they can only be transferred elastically between adjacent CrI$_3$ layers when their relative alignment is not antiparallel.  

\section{Discussion and conclusions.-} Our results provide a plausible explanation for two different experimental observations. First, the  kink in the differential spin susceptibility $\frac{d\chi}{dT}$ observed in bulk CrI$_3$ at the structural phase transition\cite{McGuireSales2015}. Second,  the antiferromagnetic interlayer coupling observed for few-layer CrI$_3$ bilayers, at low temperatures \cite{HuangXu2017,SongXu2018,KleinJarillo2018,HuangXu2018}.  Given that in these experiments the CrI$_3$ layers are either deposited on a substrate\cite{HuangXu2017} or embedded in a circuit, we conjecture that the stacking might be different than in bulk.  However, the confirmation of this hypothesis will require further experimental and computational work. 

To summarize, we have computed the interlayer exchange for CrI$_3$ bilayers in two different stacking,  that correspond to the rhombohedral and monoclinic structures observed for bulk CrI$_3$.  We find that the interlayer coupling shows a much weaker ferromagnetic character for the monoclinic than the rhombohedral phase, and eventually undergoes an antiferromagnetic transition under shear strain. We claim that this provides a possible explanation for two different experimental observations. First, the  kink observed in the differential susceptibility at the structural transition in bulk.\cite{McGuireSales2015}  Second, the fact that CrI$_3$ bilayers deposited on graphene are known to have antiferromagnetic interlayer coupling.\cite{HuangXu2017,WangMorpurgo2018,SongXu2018,KleinJarillo2018,HuangXu2018} 

\emph{Note:} 
During the  final completion  of this manuscript, we became aware of the work of four other groups\cite{JiangJi2018,SivadasXiao2018,JangHan2018} addressing the relation between interlayer exchange and stacking in CrI$_3$ using different approximations and obtaining results compatible with ours.

\emph{Acknowledgements.-} 
We acknowledge Efr{\'e}n  Navarro-Moratalla  for pointing out the different interlayer coupling for bilayer and bulk CrI$_3$. We thank Jos{\'e} Luis Lado, Francisco Rivadulla , A. H. MacDonald and  Jeil Jung  for fruitful discussions. DS thanks NanoTRAINforGrowth Cofund program at INL and the financial support from EU through the MSCA Individual Fellowship program at Radboud University. J. F.-R. acknowledge financial support from FCT for the P2020-PTDC/FIS-NAN/4662/2014, the P2020-PTDC/FIS-NAN/3668/2014 and the UTAPEXPL/NTec/0046/2017 projects, as well as Generalitat Valenciana funding Prometeo2017/139 and MINECO Spain (Grant No. MAT2016-78625-C2). CC and JFR acknowledge FEDER project NORTE-01-0145-FEDER-000019. The authors thankfully acknowledge the computer resources at Caesaraugusta and the technical support provided by the Institute for Biocomputation and Physics of Complex Systems (BIFI) (RES-QCM-2018-2-0032). Part of this work was carried out on the Dutch national e-infrastructure with the support of SURF Cooperative.

\bibliography{biblio}

\end{document}